TITLE

# Cameraless High-throughput 3D Imaging Flow Cytometry

Yuanyuan Han[1], Rui Tang[1], Yi Gu[1], Alex Ce Zhang[1], Wei Cai[2], Violet Castor[1], Sung Hwan Cho[3], William Alaynick[3], Yu-Hwa Lo[1]*

[1]Department of Electrical and Computer Engineering, University of California San Diego, La Jolla, CA, USA.

[2]Materials Science and Engineering Program, University of California San Diego, La Jolla, CA, USA.

[3]NanoCellect Biomedical Inc, San Diego, CA, USA

*e-mail: ylo@ucsd.edu

ABSTRACT

Increasing demand for understanding the vast heterogeneity of cellular phenotypes has driven the development of imaging flow cytometry (IFC), that combines features of flow cytometry with fluorescence and bright field microscopy. IFC combines the throughput and statistical advantage of flow cytometry with the ability to discretely measure events based on a real or computational image, as well as conventional flow cytometry metrics. A limitation of existing IFC systems is that, regardless of detection methodology, only two-dimensional (2D) cell images are obtained. Without tomographic three-dimensional (3D) resolution the projection problem remains: collapsing 3D information onto a 2D image, limiting the reliability of spot counting or co-localization crucial to cell phenotyping. Here we present a solution to the projection problem: three-dimensional imaging flow cytometry (3D-IFC), a high-throughput 3D cell imager based on optical sectioning microscopy. We combine orthogonal light-sheet scanning illumination with our previous spatiotemporal transformation detection to produce 3D cell image reconstruction from a cameraless single-pixel photodetector readout. We further demonstrate this capability by co-capturing 3D fluorescence and label-free side-scattering images of single cells in flow at a velocity of 0.2 m s$^{-1}$, corresponding to a throughput of approximately 500 cells per second with 60,000 voxels (resized subsequently to 10$^6$ voxels) for each cell image at a resolution of less than 1 micron in X, Y, and Z dimensions. Improved high-throughput imaging tools are needed to phenotype-genotype recognized heterogeneity in the fields of immunology, oncology, cell- and gene- therapy, and drug discovery.

TEXT

A central challenge of biology is to correlate the phenotype of heterogeneous individuals in a population to their genotype in order to understand the extent to which they conform to the observed population behavior or stand out as exceptions that drive disease or the ability to become threats to health[1,2,3,4]. While optical microscopy is a cornerstone method to study the morphology and molecular composition of biological specimens, flow cytometry is a gold standard for quantitative high-throughput single-cell characterization in numerous biomedical applications[5,6]. Recognizing the need to merge these two powerful platforms, several groups have proposed techniques for imaging flow cytometry (IFC)[7]. IFC simultaneously produces ensemble-averaged measurements and high-content spatial metrics from individual cells in a large population of cells, without perturbation due to experiment condition change. An important limitation of existing IFC systems is that, regardless of the optical detection method and computation algorithm is used, only 2D cell images can be obtained[8,9,10]. The absence of 3D tomography results in occlusion of objects, blurring by focal depth, loss of z-axis spatial resolution, and artifacts due to projection of a 3D cell into a 2D image. For example, with 2D microscopic imaging, if a fluorescent probe is observed at the center of a cell, its location (e.g. membrane, cytosol, nucleus) is ambiguous. For a range of applications, such as internalization measurements, probe co-localization, and spot counting, relative to 2D imaging that is dependent on the cellular orientation to the imaging plane, 3D images provide more complete and accurate phenotyping of cell and organelle morphology, as well as nucleic acid and protein localization to support biological insights[11]. Unfortunately, rapid and continuous 3D image acquisition for single cells in flow has not been available.

Here we present high-throughput three-dimensional imaging flow cytometry (3D-IFC) based on optical sectioning microscopy[12]. This combination of light-sheet scanning illumination technique and spatial-temporal transformation detection technique enables fluorescent and label-free 3D cell image reconstruction from single-element photodetector readout without a camera[13][14]. Building upon the speed and sensitivity benefits of photomultiplier tube (PMT), the 3D-IFC uses multiple scanning techniques to add spatial information in a fairly conventional flow cytometry architecture. 3D imaging is achieved by laser scanning across the first (z-) axis, the cell translating by flow across the second (y-) axis, and the use of multiple pinholes arranged along the third (x-) axis to produce fluorescent and label-free information from 60,000 voxels per cell. By precisely mapping time to space, photodetector readout at one timepoint corresponds to one voxel in a 3D space. Here we demonstrate 3D-IFC of fluorescence and 90-degree label-free side-scattering (SSC) imaging of single cells in flow at a velocity of 0.2 m s$^{-1}$, corresponding to a throughput of approximately 500 cells per second.

A schematic of the 3D-IFC system is shown in Figure 1a. In the 3D-IFC system, suspended cells form 2D hydrodynamically focused single file in a quartz flow cell with a square cross section (Extended Data Figure 1a)[15]. Laser excitation is via a light-sheet (x-y plane) with a diffraction limited beam waist and a height of 200 to 400 μm, scanning in z-direction at a very high rate (200 kHz). When a cell flowing through the whole optical interrogation at 0.2 m s$^{-1}$, a pixelated field of view is represented by a 3D space with X by Y by Z voxels as shown in Figure 1c. A pinhole array on the spatial filter is aligned at a tilting angle, $\vartheta$, to the flow stream, so the pinhole array also steps along x-direction. In this manner, each pinhole allows light from voxels with a distinct x-index to pass to PMT detector (see Methods for mask details). The imaging process begins when a flowing cell appears at the first pinhole of the spatial filter. During the first light-sheet scanning period (5 μs), light intensity of voxels $z_{1-Z}$ with $x_1 y_1$ index is collected. As the cell flows downstream in $y$ to the next position, $x_1 y_2$, the corresponding $z_{1-Z}$ voxels are produced. This is repeated until cell completely passes pinhole 1 when the whole 2D yz-slice at $x_1$ is imaged. As the cell travels farther downstream in $y$, it reaches the following pinholes and yz-slices of at $x_2$ to $x_X$ are recorded. A detailed description and mathematical formulation are included in the Methods.

To precisely measure the speed of each cell for image reconstruction by the spatiotemporal transformation, pairs of slits upstream and downstream of the pinholes are added to the optical spatial filter (Extended Data Figure 1b). The measured rate of the cell moving through the detection zone, and the known frequency of the light-sheet scanning rate ensures each voxel in a 3D-IFC image has a distinct time-domain value, and all voxels can be discreetly captured in time from single-element PMT. 3D cell images are then reconstructed from the time-domain signal. Figure 1d-f shows the typical output of the 3D-IFC system. The detected time-domain multi-parametric signals (multi-color fluorescence, FL1 and FL2, and side-scattering, SSC, light intensities) are synchronized with the reference output of the tuning voltage of the AOD driver, which denotes the z-position of the detected voxel. At a 200 kHz scanning rate, a one-dimensional (1D) intensity profile in z-axis is recovered from the time-domain signal within a time period of 5 μs (Fig. 1d). During the ~100 μs of cell travel between pinholes, ~20 periods of laser scanning are performed, and an yz-plane 2D image array is recovered from the PMT readout (Fig. 1e). As the cell travels through the entire interrogation area, a stack of 2D yz-plane images are recovered, and the final 3D image is reconstructed, with the ordinal pinhole number indicating the voxel's x-position. In the example shown in Figure 1f, a 10-pinhole spatial filter produces 10 2D yz-plane images and a signal length of 1 ms, corresponding to a throughput of 500 cells per second. The bandwidth of PMT and digitizer, 150 MHz and 125 MHz, respectively, in this implementation, supports the throughput at 60,000 voxels per 3D cell image. Each 3D cell image in a 3D space of 20 μm by 20 μm by 20 μm, is resized to 100 by 100 by 100 pixels for 3D image analysis and quantitative measurements.

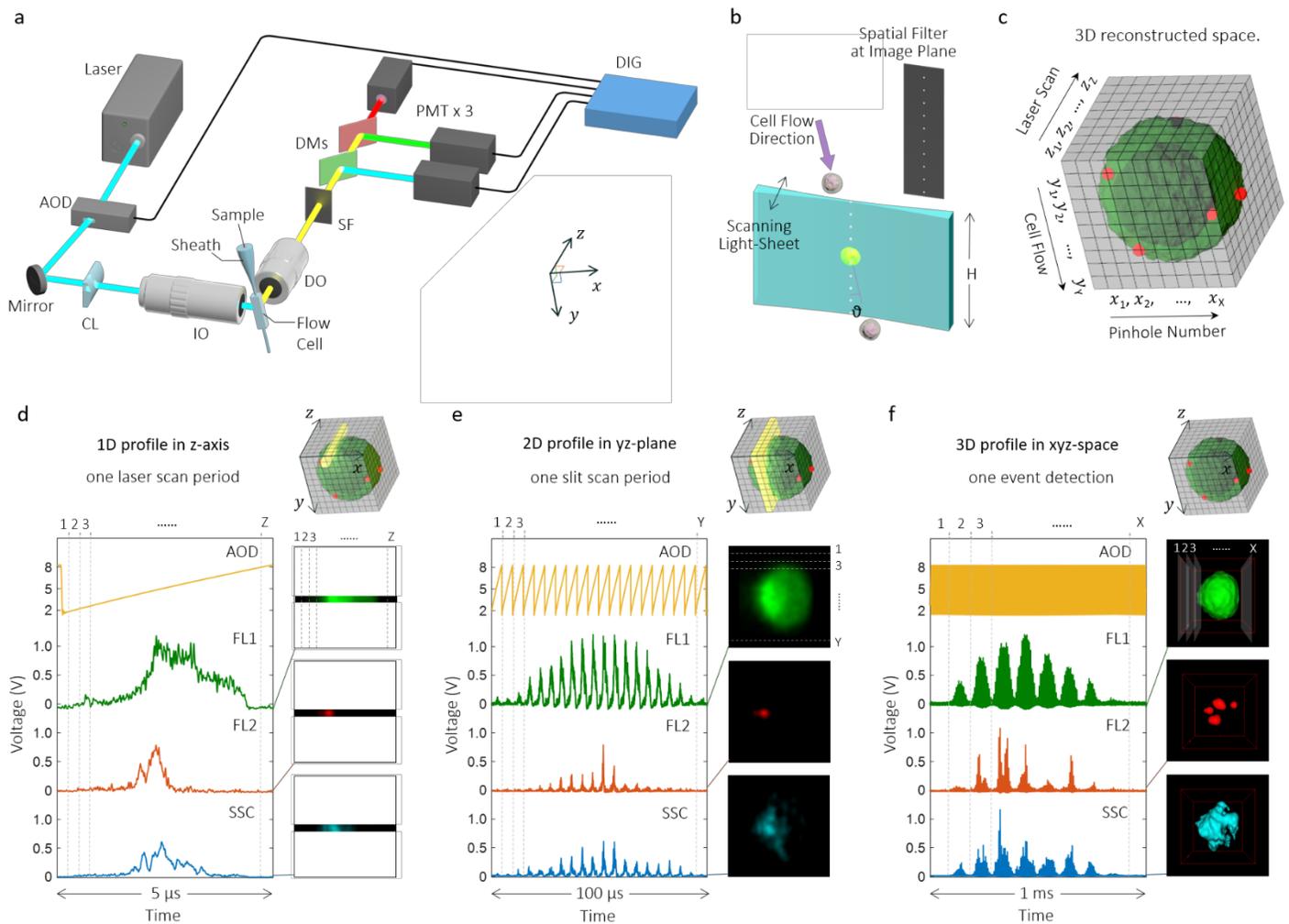

**Figure 1. Implementation of the 3D-IFC System and Demonstration of Time to 3D-space Mapping.**

**(a)** Schematic diagram of the 3D-IFC system. AOD, acousto-optic deflector; CL, cylindrical lens; IO, 50X/0.55 illumination objective lens; DO, 10X/0.28 detection objective lens; SF, spatial filter; DMs, dichroic mirrors; PMT, photomultiplier tube; DIG, 125 MS s$^{-1}$ digitizer. The AOD and CL produce a scanning light-sheet. Sample is 2D hydrodynamically focused by sheath before entering the square cross section quartz flow cell. **(b)** Optical interrogation area. H, height of the light-sheet; $\vartheta$, tilt angle between flow (y-axis) and vertical line. Illumination light-sheet propagates horizontally and scans in z-axis, sample flows in y-axis, x is the orthogonal axis. The spatial filter at the image plane uses pinholes to produce line scans across the x-axis. **(c)** 3D reconstructed space. The resolution on the X-axis is determined by the number of pinholes (pixelated field of view in x-direction); resolution of Y by the distance between two slits (pixelated field of view in y-direction); and resolution of Z by the light-sheet scanning range (pixelated field of view in z-direction). **(d)** One light-sheet scan period produces 1D light intensity profile in z-axis. The PMT voltage readout of one timepoint corresponds to the light intensity of one voxel in z-axis. **(e)** While object travels along y-axis, multiple scans produce a 2D profile in yz-plane within one pinhole scan period. Each section—separated by dotted lines—corresponds to the light intensity of one row in the 2D image stack. **(f)** When object completely passes through the spatial filter covering area, the time-domain signal contains the complete information of the 3D profile in xyz-space. Each section corresponds to one 2D image slice. AOD, tuning voltage of the AOD driver; FL1, PMT readout of fluorescence detection channel 1; FL2, PMT readout of fluorescence detection channel 2; SSC, PMT readout of side-scattering light detection channel.

To demonstrate the cellular imaging capability of 3D-IFC system, we imaged suspended single cells at a flow speed of 0.2 m s$^{-1}$. Figures 2-4 show two-color fluorescence and unlabeled dark-field (side-scattering) 3D images of mammalian cells. In Figure 2, HEK293 cells were stained with an intracellular carboxyfluorescein dye (CFSE) and bound with a random number of 1 μm fluorescent carboxylate-modified polystyrene beads. Figure 2b shows that while 2D bead images overlapped, the 3D-IFC resolved the exact number of particles from the reconstructed 3D images, which is crucial for localizing and co-localizing features[16,17]. In a flow system, cells and their internal structures are orientated at random, as a result 2D images may be from an unfavorable viewing perspective. Multiple perspectives can be achieved via 3D-IFC cell tomography to provide improved relative position relationships and spot counting results, which favors both machine vision and human visualization. Using the side-scatter dark-field imaging mode, the 3D-IFC generates a 3D spatial distribution of scattered light. It is known that refractive index variations will scatter light when the object is illuminated by visible light, and the size and refractive index distributions of the scattering regions determine the distributions of the scattered light. The 3D SSC images represent the spatial distribution of those refractive index (n) variations among the fluid (PBS, n~1.335), the cells (n~1.3-1.6) and the polystyrene beads (n~1.6)[18,19]. As shown in Figure 2c, intensity-based low-pass filtered SSC image indicates cell volume, and high-pass filtered SSC image correlates with the fluorescent image of beads.

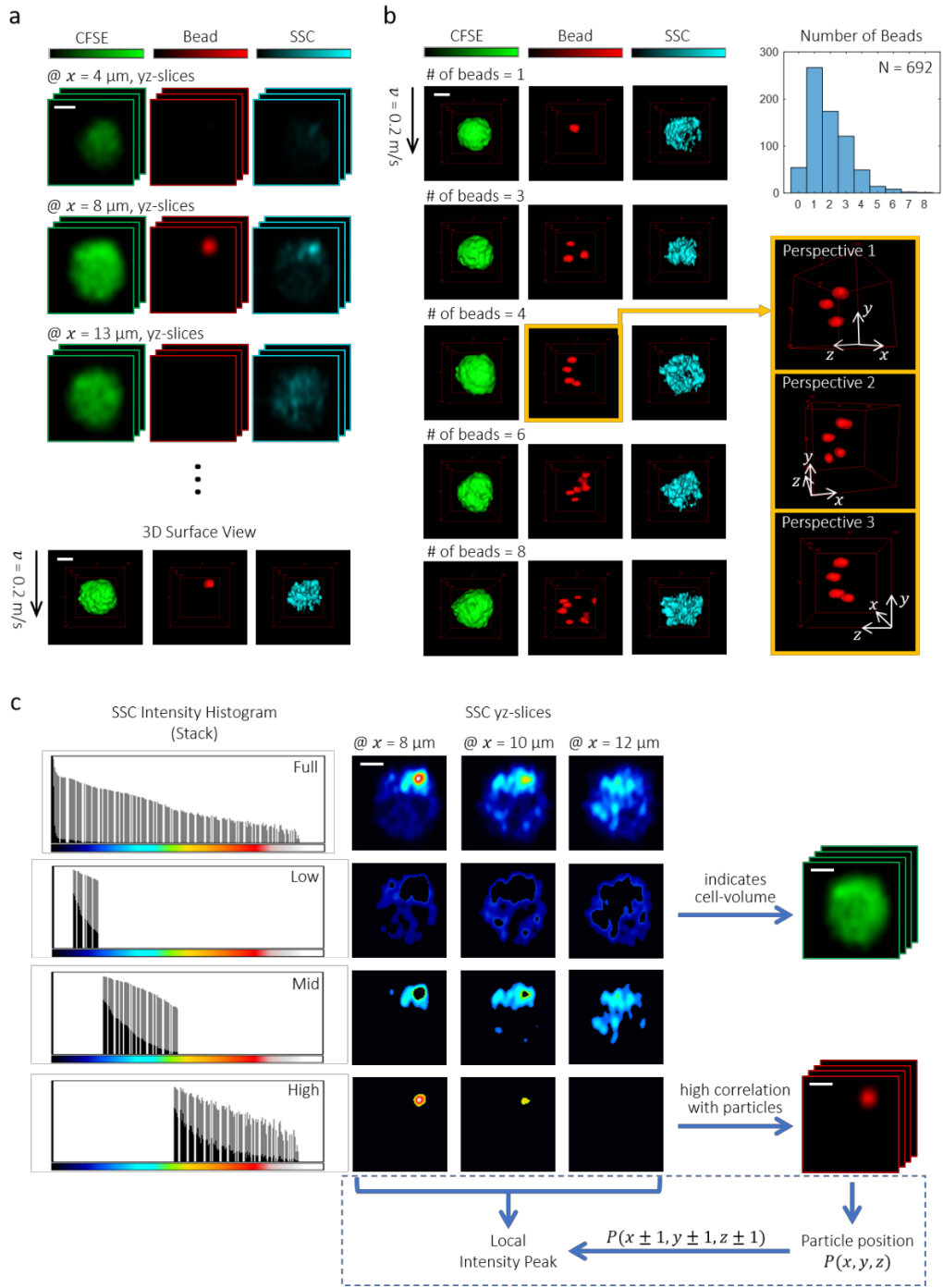

**Figure 2. Cells and Beads Imaged by the 3D-IFC.**

CFSE-stained HEK-293 cells bound with 1 μm fluorescent beads. **(a)** Recovered 2D yz-plane images and the assembled 3D surface-rendered view of CFSE fluorescence, bead, and SSC (bottom row). **(b)** Representative 3D images of cells bound with beads and histogram detection events. The explicit relative position relationship in 3D space indicates that the particle counting in the 3D-IFC is independent of cell orientation. In the example of cell bound with 4 beads, occlusion in specific perspective is a likely source of error for particle counting with 2D images. **(c)** Intensity-based processing of 3D SSC images. Left column: intensity histograms of 3D SSC image of the cell shown in Figure 3a. $P(x, y, z)$ is the position of 1 μm size bead determined using 3D fluorescent image; within each bead position's $\pm 1$ μm area, the local intensity peak in 3D SSC image can be found. Scale bars, 5 μm. Flow speed 0.2 m/s. CFSE, intracellular carboxyfluorescein dye, Ex/Em: 488/517; Bead, carboxylate-modified fluorescent microspheres, Ex/Em: 488/645; SSC, 90-degree side-scattering.

.

When ionizing radiation or cytotoxic chemical agents cause DNA damage in the form of double stranded breaks (DSBs), the phosphorylated protein gamma-H2AX (γH2AX) quickly forms foci at DSBs in a 1:1 manner[20]. With anti-γH2AX immunolabeling, foci reflect DNA damage and ability for DNA repair[21]. 2D imaging techniques and manual quantification are used today, but counting foci from 2D images is labor intensive and unreliable due to perspective dependence[22]. To evaluate foci counting of the 3D-IFC system, we imaged immunolabeled γH2AX foci in CMK3 cells (a glioblastoma multiforme cell line) after 6 Gy of gamma-irradiation. Representative cell images are shown in Figure 3. The data show that the number of γH2AX foci is unrelated to the fluorescence intensity, thus intensity-based measurements with conventional flow cytometry metrics poorly reflects DNA damage. In contrast, the 3D-IFC system enabled accurate and rapid analysis of the number of γH2AX-positive foci.

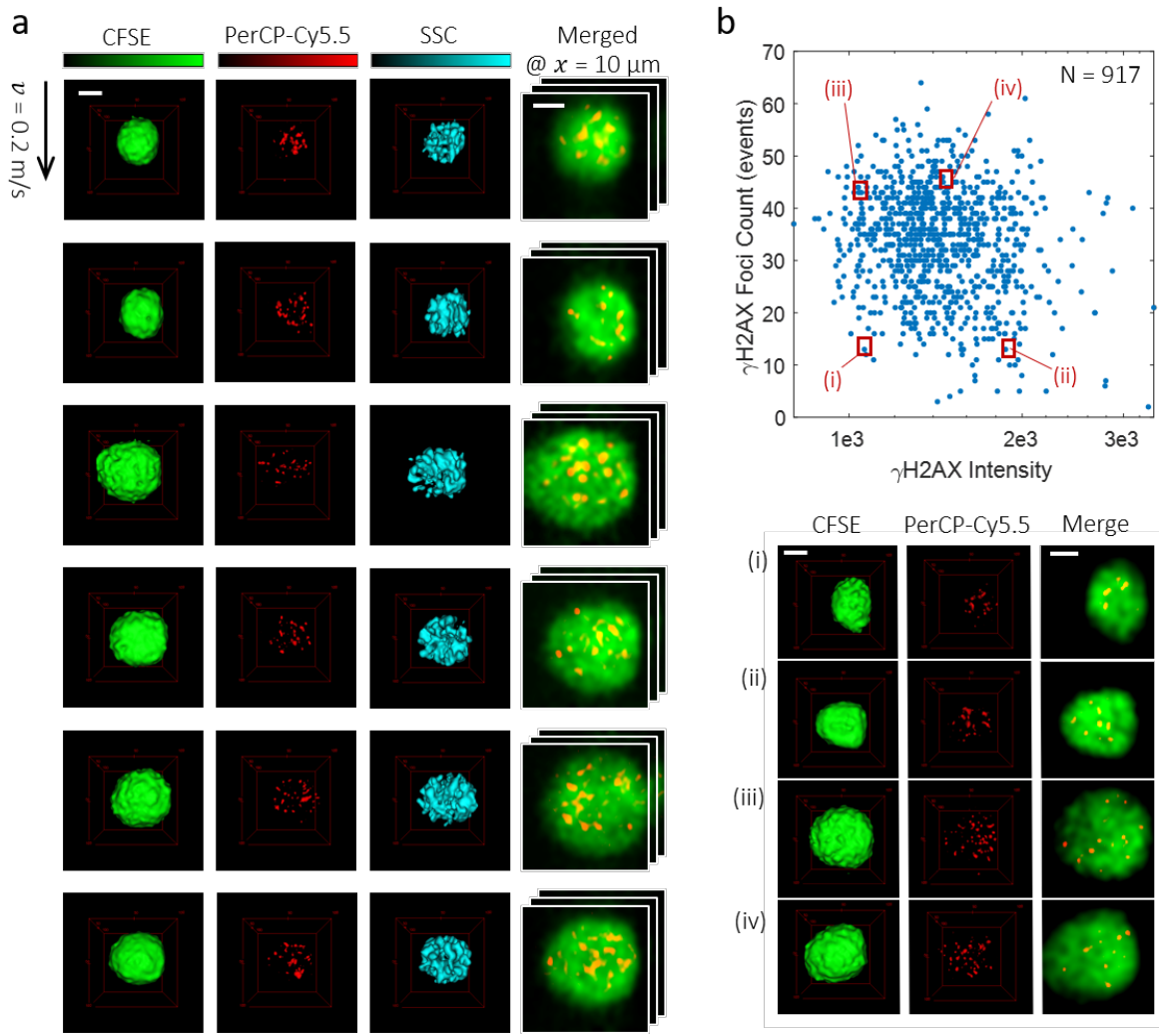

**Figure 3. Fluorescent γH2AX foci Imaged by the 3D-IFC.**

(a) Representative 3D images of irradiation damaged glioblastoma CMK3 cells stained with CFSE and γH2AX antibody conjugated PerCP/Cy5.5; and their two-color fluorescence 2D yz-plane merged image slices at x = 10 μm. The high quality of the 3D images shows that the 3D-IFC is suitable for DNA-damage foci related study. (b) Scatterplot of 917 detection events in the γH2AX intensity and foci count together with images of the cells within the marked regions (i)–(iv) in the scatterplot. The data show that foci count is unrelated to the fluorescence intensity from labelled γH2AX, thus intensity-based measurements with conventional flow cytometry metrics are unable to evaluate the extent of DNA damage. Scale bars, 5 μm. The flow speed 0.2 m/s. CFSE, intracellular carboxyfluorescein dye, Ex/Em: 488/517; PerCP-Cy5.5: DNA-damage antibody conjugated dye, Ex/Em: 490/677.

Peripheral blood leukocyte morphology is important clinical diagnostic and prognostic measure for acute and longitudinal evaluations[23]. Figure 4a shows representative 3D-IFC imaging of leukocytes in three imaging modes (Extended Data Figure

2): 2D transmission image (a 3D-IFC parameter, not discussed) and 3D fluorescence and side-scattering images. The leukocyte SSC signal not only indicates nuclear granularity but also provides cell volumetry obtained from low intensity (low refractive index) SSC imaging. Here two or more intensity bands of 3D SSC signal can be used to generate two or more images that reflect their corresponding refractive indices, such as the cytosol and nucleus, whereas conventional flow SSC intensity signal is dominated by contributions of the nucleus. As shown in Figure 4b-c, we can use 3D SSC image to produce cell volume that matches the transmission and fluorescence results.

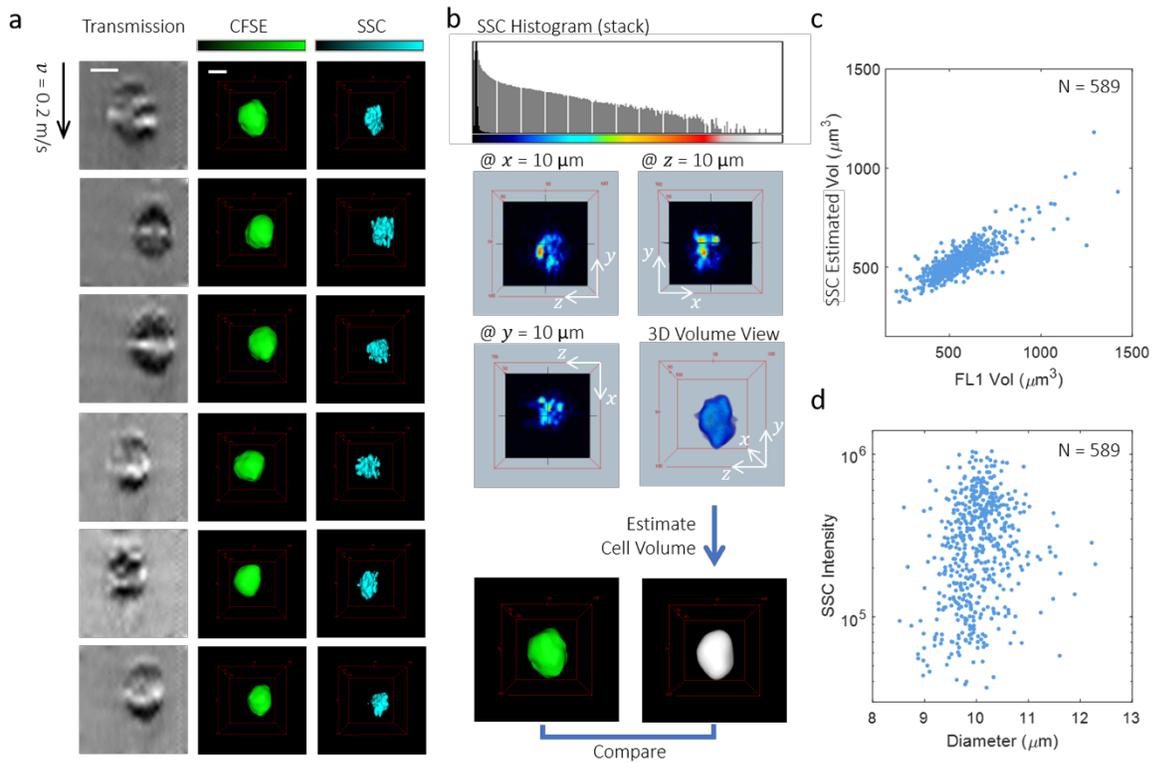

**Figure 4. Leukocytes Imaged by the 3D-IFC.**

(a) Representative 2D transmission images (left column) and 3D images of leukocytes. (b) Intensity histogram of SSC signal of the first cell shown in (a) and its 3D profile in Orthoslice view and Volume view. Cell volume estimation based on 3D SSC image matches the fluorescence volume (see Supplementary Data). (c) Scatterplot of 589 detection events in the CFSE (FL1) volume and SSC-based estimated cell volume. (d) Scatterplot in the cell diameter calculated from SSC-based estimated cell volume and SSC intensity. Scale bars, 5 μm. Flow speed 0.2 m/s.

Through a confluence of several technological discoveries, we have demonstrated that 3D-IFC provides an unprecedented access to high-throughput 3D image capture and analysis. This implementation of 3D-IFC features a cell flow speed of 0.2 m s$^{-1}$, a spatial resolution of 1 μm in 3 axes, a field of view of 20 μm in all axes, and a throughput of 500 cells per second. The design has the flexibility to directly increase the spatial resolution, field of view and 3D image capture rate through the change of spatial filter and the use of higher flow rate. Its information-rich 3D dark-field (side-scattering) image detection, coupled with 2D transmission image, offers possibilities for label-free assays. Potential applications, such as asymmetric division of T-cells into effector and memory cells[24], the secretory pathway of B cells, phenotype drug discovery[25,26], protein or receptor translocations[27,28], tracking of organelle formation or trafficking[29], and chromosome structural aberrations[30], where 3D orientation and polarity are important, could greatly benefit from this 3D-IFC technology, especially in future integrations with image-based sorting functionality[31,32].

## ACKNOWLEDGEMENTS

This work was performed in part at the San Diego Nanotechnology Infrastructure (SDNI) of UCSD, a member of the National Nanotechnology Coordinated Infrastructure (NNCI), which is supported by the National Science Foundation (Grant ECCS-1542148). Research reported in this publication was supported by the National Institutes of Health under award number 1R43DA045460. The content is solely the responsibility of the authors and does not necessarily represent the official views of the National Institutes of Health.

## AUTHOR CONTRIBUTIONS

Y.H. and Y.L. conceived the concept of the 3D-IFC and designed the research. Y.H. wrote the manuscript. Y.H. constructed the 3D-IFC system. Y.H. and R.T. performed experiments and data/image processing. W.A. suggested biological application selection and participated in writing the manuscript. Y.G., A.C.Z., W.C., V.C., S.H.C. provided assistance to experiments. Y.L. supervised the project. All authors discussed results and commented on the manuscript.

## COMPETING INTERESTS

Yu-Hwa Lo has an equity interest in NanoCellect Biomedical, Inc. as a co-founder and a member of the company's Scientific Advisory Board. NanoCellect may potentially benefit from the results of this research.

## ADDITIONAL INFORMATION

**Extended data** is available for this paper.

**Correspondence and requests for materials** should be addressed to Y.L.

## METHODS

### Microfluidic System

The microfluidic system (Extended Data Figure 1a) continuously introduces suspended cells to the optical interrogation area. Since the position of the cells in the cross-section of the microchannel is aligned with the optical field of view in all the x-, y- and z-directions, a reproducible and stable flow speed for cells is obtained by tightly focusing the cells at the center of the square cross-section. Suspended cells in a sample injected by a syringe pump are hydrodynamically focused into a single stream. A sheath flow is used to confine flowing cells in both x- and z-direction. An air pressure pump together with a liquid flow meter are used to provide stable sheath flow. At the junction of sheath flow and sample flow, tubing ends are specially tapered to keep symmetric flow. The flow rate ratio between the sample and sheath is precisely controlled to be 100:1 to ensure particles flowing at a high speed and within the optical field of view.

### 3D Image Detection and Construction

The optical system is arranged in a light sheet fluorescence microscopy configuration (Figure 1a), which only illuminates a specimen in a single plane at a time whilst the signal is detected in a perpendicular (z-) direction. An acousto-optic deflector (AOD) is producing the high-speed, 200 kHz in this implementation, scanning in z-direction. A single-pixel photodetector detects fluorescence or scattering light passed the spatial filter in each channel, and an image is reconstructed from the time-domain output. The pinholes on the spatial filter is arranged vertically. The flow (y-) direction is tilted to create an angle $\vartheta$ (Fig. 1b), which is determined by the field of view in x-direction $D_x$, the field of view in y-direction $D_y$ and number of pinholes $X$, $\vartheta = \tan^{-1}(D_x/(X \cdot D_y))$. Also, because the z-direction illumination light-sheet scans over a range that is larger than the cell size, and at a speed that is much higher, typically more than 20 times higher than the cell travelling speed in y-direction, when cell passing one pinhole, the light intensity of one yz-plane image slice is recorded; when cell passing the next pinhole, another yz-plane image stack is acquired. Combining the scanning light-sheet, the cell's flow motion, and the spatial filter, the detector only detects the fluorescence of an individual voxel in the cell at a time, which allows one-to-one time to space mapping. The concept can be mathematically formulated in the following equations. The measured PMT signal $S(t)$ can be expressed as

$$S(t) = \iiint dxdydz \left\{ \iint dx'dy' \, C(x', y' - v_c t, z) I(z, t) \, psf(x - x', y - y') dx'dy' \right\} F(Mx, My)$$
$$= \iiint dxdydz \, \{C(x, y - v_c t, z) I(z, t)\} \otimes psf(x, y, z)\} \cdot F(Mx, My)$$

, where $C(x, y, z)$ is the 3D cell fluorescence or scattering light intensity profile, $I(z, t)$ is the light-sheet illumination, $F(x, y)$ is the characteristic function of the spatial filter, $v_c$ is the cell flowing speed, $M$ is the magnification factor of the detection system.

The acoustic frequency sent to the acoustic transducer in the AOD is varied to deflect the beam to create illumination at different z-position. The tuning voltage that produce continuous change of acoustic frequency can be generated by various types of waveforms, such as sinusoidal, triangle, etc. For the most laser power efficient, the tuning voltage is set to be changing in a sawtooth manner, so the position of the light sheet in z-direction $z_0(t)$ can be described as

$$z_0\big(t \in (nT, (n+1)T)\big) = v_i(t - nT), \quad n = 0,1,2,\ldots$$

, where $T$ is the light-sheet illumination scanning period, $v_i$ is the scanning speed in z-direction.

By using the cylindrical lens, the laser is diverged to form a light-sheet with a height in y-direction of 200-400 μm. The scanning light-sheet illumination $I(z, t)$ can be described as Gaussian beam:

$$I(z, t) = k \cdot e^{-\frac{(z - z_0(t))^2}{\sigma^2}}$$

With oversampling PMT signal readouts, the spatial resolution in z-direction is diffraction limited. The Gaussian beam waist is measured 0.73 μm and is approximated as a delta function for simpler calculation:

$$I(z,t) \approx k \cdot \delta(z - z_0(t))$$

Two examples of the spatial filter are shown in Extended Data Figure 1b. Putting the slits used for speed detection aside, the characteristic function $F(x,y)$ of the spatial filter is designed to be

$$F(x,y) = \sum_{q=1}^{N} \delta(x - x_q) \cdot \delta(y - y_q)$$

, where $q = 1, 2, \ldots, N$ is the number of pinholes on the spatial filter. The size of the pinhole, together with the NA of detection objective lens, the cell flowing speed, and signal sampling rate determine the spatial resolution in x- and y-directions. The spatial filters are fabricated using electron beam lithography and the blackout area is made of chromium with a thickness of 250 nm.

With the approximations above, when $y_{q+1} - y_q > cell\ size$, and cell projection is overlapped with $j$-th pinhole,

$$S(t) = \int_{x,y} C(x, y - Mv_2 t, z_0(t)) \delta(x - x_q) \cdot \delta(y - y_q) dx dy$$

$$= C(x_j, y_j - Mv_c t, z_0(t))$$

Presuming cell is within the depth of field, and the system PSF at multiple $z$ is not considered. The light intensity signal detected by PMT (Hamamatsu) is first amplified by an amplifier (Hamamatsu) with a bandwidth from DC to 150 MHz, and then digitized by a digitizer (ADVANTECH) with a maximum sampling rate of 125 MS/s per channel.

Consequently, this approach maps the 3D cell image into the time-domain light intensity on a one-to-one basis. The 3D image construction algorithm is realized in MATLAB according to the equation above. Due to slight variance in flowing speed $v_c$ of cells, the original size of the 3D image of each cell can be slightly different. The original 3D image is then resized to 100 x 100 x 100 pixels. 3D image batch processing is performed in ImageJ.

### TESTED SAMPLE DETAILS

### Cell with Fluorescent Beads

The human embryonic kidney 293 (HEK-293) cells were cultured with complete culture media (DMEM, 10% Fetal Bovine Serum, 1% Penicillin Streptomycin) to 90% confluency in 10 cm petri dish. After 100X dilution of the 1.0 μm fluorescent beads (Ex/Em: 488/645 nm, T8883, ThermoFisher) from the stock solution (2% solids), 100 μL of the diluted solution was mixed with 10 mL fresh cell culturing media and added to cell culturing plate. After continuous culturing for 10 hours, the HEK-293 cells were harvested and stained with CellTrace CFSE cell proliferation kit (Ex/Em: 492/517 nm, C34554, ThermoFisher) at a working concentration of 20 μM. After the staining process, cells were fixed by 4 % paraformaldehyde, washed and resuspended in 1X phosphate buffered saline (PBS). Before every imaging experiment, the cell suspension was diluted in PBS to a concentration of 1000 cells/μL.

### CFSE Staining

The HEK-293 cells were first cultured with complete culture media to 98% confluency in 10 cm petri dish, and then were harvested and resuspended to a concentration of $1 \times 10^6$ cells/mL in 1X PBS. The CFSE Cell Proliferation Kit (Ex/Em 492/517nm, C34554, Thermo Fisher) were added to the cell suspension at a working concentration of 20 μM. After incubating the cells at 37°C for 30 minutes, fresh culture media (DMEM) were used to quench the staining process and the HEK-293 cells were washed by 1X PBS and fixed by 4 % paraformaldehyde. The fixed cells were washed and resuspended in 1X PBS. Protocol is also applied to stain CMK3 cells and human blood leukocytes.

**CMK3 Cell Irradiation Treatment and Immunostaining**

The human glioblastoma CMK3 cells were cultured with complete culture media (DMEM-F12, 2% B27 supplement, 1% Penicillin Streptomycin, 1% Glutamax, 100 µg/L EGF, 100 µg/L FGF, 0.24% Heparin) in culture plates. The cells were harvested and resuspended to a concentration of $1x10^6$ cells/mL in 1X PBS, and then stained with CFSE. To induce DNA double-strand breaks (DSB), CMK3 cells are treated with 6 Gy irradiation by cesium source irradiator. The treated cells were washed once with 1X PBS and fixed with 1% paraformaldehyde 30 minutes post irradiation. The fixed cells were washed with PBS twice. Then 70% ethanol was added to the cells and the cells were incubated on ice for 1 hour. After ethanol treatment, cells were washed with PBS twice and incubated in 1% TritonX-100 at room temperature for 10 minutes. Then, cells were washed with PBS once and incubated in 5% Bovine Serum Albumin (BSA) in PBS for 30 minutes at room temperature on shaker. Cells were then washed with PBS once and incubated in anti-phospho-histone H2A.X (Ser139) Antibody, clone JBW301 at 1:300 dilution on ice on shaker for 1 hour. After the primary antibody treatment, cells were washed twice with 5% BSA and incubated in PerCP/Cy5.5 anti-mouse IgG1 Antibody at 1:100 dilution on ice on shaker for 1 hour. At last, the stained cells were washed twice with 5% BSA and resuspended in 1:3 diluted stabilizing fixative buffer in MilliQ water. Before every imaging experiment, the cells ware diluted in PBS to a concentration of 500 cells/µL.

**Leukocytes Separation and staining**

The fresh human whole blood was harvested in EDTA tube from San Diego Blood Bank. The red blood cells in the whole blood were first lysed by RBC lysis buffer (00-4300-54, Invitrogen) and the leukocytes population was harvested by soft centrifuge after the lysing process. The harvested leukocytes were then washed and resuspended to a concentration of $1x10^6$ cells/mL in 1X PBS. The resuspended leukocytes were stained with CFSE. After the staining process, the cells were fixed by 4 % paraformaldehyde, washed and resuspended in 1X PBS. Before every imaging experiment, the cell suspension was diluted in PBS to a concentration of 500 cells/µL.

**DATA AVAILABILITY**

The datasets generated and analyzed in this study are available from the corresponding author on reasonable request.

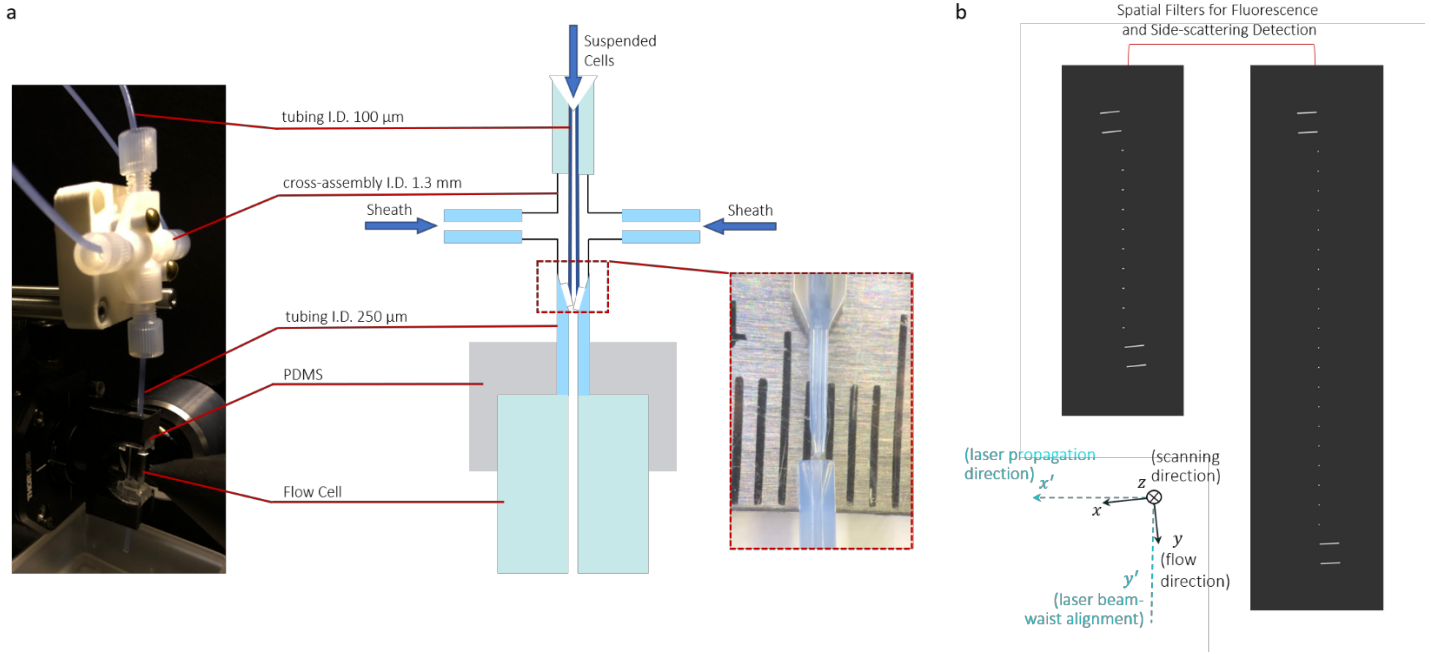

**Extended Data Figure 1. Spatial Filters and Microfluidic System in the 3D-IFC. (a)** Microfluidic system in the 3D-IFC together with picture of specially engineered tapered tubing. The flow through channel in the quartz flow cell has a cross-section dimension of 250 μm by 250 μm and a length of 20 mm. **(b)** Two examples of spatial filters placed at image plane. The top two and bottom two long slits with dimensions of 10 μm by 200 μm are for speed detection. The other pinholes on the spatial filter are 10 μm by 20 μm (left) and 10 μm by 10 μm (right), for 3D image capturing with pixel size of 2 μm and 1 μm in x-direction, respectively. The arrangement of pinholes aligns the laser beam-waist.

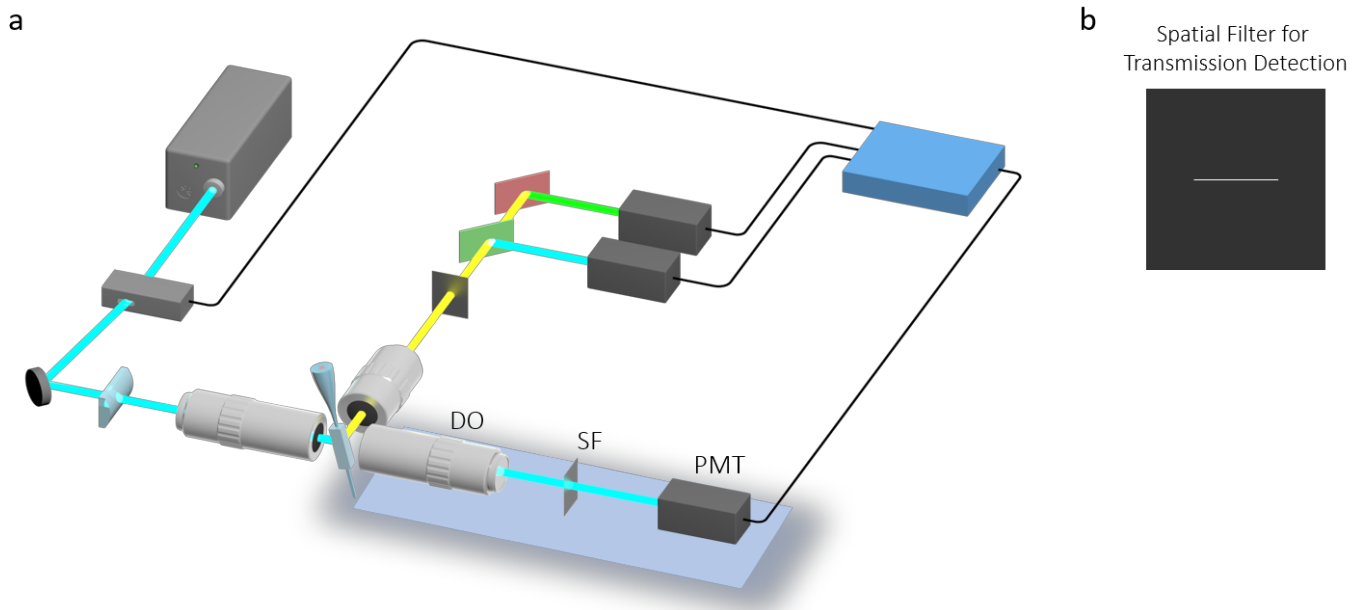

**Extended Data Figure 2. Implementation of 2D Transmission Imaging Mode.**

**(a)** Schematic diagram of the 3D-IFC with transmission imaging mode. Shadowed part is added for transmission image detection. DO, 50X/0.55 detection objective lens; SF, spatial filter; PMT, photomultiplier tube. **(b)** Spatial filter for transmission detection. Single slit with dimensions of 50 μm by 2 mm.